\documentclass[english]{elsarticle}
\usepackage{geometry}
\usepackage{color}
\usepackage{float}
\usepackage{mathrsfs}
\usepackage{mathtools}
\usepackage{amsthm}
\usepackage{graphicx}
\usepackage[final]{changes}
\usepackage[toc,page]{appendix}
\usepackage{multirow}

\numberwithin{table}{section}
\numberwithin{figure}{section}

\makeatletter

\providecommand{\tabularnewline}{\\}

\theoremstyle{plain}

\theoremstyle{remark}

\usepackage{babel}

\@ifundefined{showcaptionsetup}{}{%
 \PassOptionsToPackage{caption=false}{subfig}}
\usepackage{subfig}
\makeatother

\usepackage{babel}
\providecommand{\remarkname}{Remark}
\providecommand{\theoremname}{Theorem}

\usepackage[utf8]{inputenc}

\usepackage{graphicx}
\usepackage{multirow}
\usepackage{amssymb}
\usepackage{float}
\usepackage{amsmath}
\usepackage{xcolor}
\usepackage{natbib}
\usepackage{setspace}

\usepackage{geometry}
\usepackage{mathrsfs}
\renewcommand{\arraystretch}{1.8}

\makeatother

\begin{document}

\title{Taxonomy of Cohesion Coefficients\\ for Weighted and Directed Multilayer Networks}

	\author[Bic1]{Paolo Bartesaghi}
		\author[Catt]{Gian Paolo Clemente}
		\author[Bic1]{Rosanna Grassi\corref{cor1}}
		
		\cortext[cor1]{\emph{Corresponding author. email: rosanna.grassi@unimib.it}} 
		\address[Bic1]{University of Milano - Bicocca, Via Bicocca degli Arcimboldi 8, 20126 Milano, Italy, email: paolo.bartesaghi@unimib.it; rosanna.grassi@unimib.it}
		\address[Catt]{Universit\`{a} Cattolica del Sacro Cuore di Milano, Largo Gemelli 1, 20123 Milano, Italy email: gianpaolo.clemente@unicatt.it}




	
	
\begin{abstract}
Clustering and closure coefficients are among the most widely applied indicators in the description of the topological structure of a network. Many distinct definitions have been proposed over time, particularly in the case of weighted networks, where the choice of the weight attributed to the triangles is a crucial aspect. In the present work, in the framework of weighted directed multilayer networks, we extend the classical clustering and closure coefficients through the introduction of the clumping coefficient, which generalizes them to incomplete triangles of any type.
We then organize the class of these coefficients in a systematic taxonomy in the more general context of weighted directed multilayer networks. Such cohesion coefficients have also been adapted to the different scales that characterize a multilayer network, in order to grasp their structure from different perspectives. We also show how the tensor formalism allows incorporating the new definitions, as well as all those existing in the literature, in a single unified writing, in such a way that a suitable choice of the involved adjacency tensors allows obtaining each of them. Finally, through some applications to simulated networks, we show the effectiveness of the proposed coefficients in capturing different peculiarities of the network structure on different scales.
\end{abstract}

\begin{keyword}
Multiplex Networks \sep Clustering Coefficient \sep  Clumping Coefficient \sep Tensors
\end{keyword}

\maketitle

\section{Introduction}
\label{Introduction}

In network analysis, a wide range of problems calls for an appropriate definition of the \textit{cohesion level} among nodes. Generally speaking, the cohesion in a graph measures to which extent nodes are connected to each other. Numerous strategies have been proposed over time to address this point \cite{Moody2015}.

In all of them, a first distinction concerns the difference between local measures, i.e. related to the neighborhood of a single node, and global measures i.e. referred to the entire network. In an attempt to extend the cohesion measures from single layer to multilayer networks, a first issue concerns the multiple perspectives from which it is possible to observe the network. In particular, as will be seen, it is possible to identify two intermediate levels at which cohesion measures, that are neither strictly local, nor strictly global, can be provided. 
 
Undoubtedly, one of the most famous among these indicators is the family of local and global \textit{clustering coefficients}. They were introduced in the sociological context where it was observed the tendency of social networks to form tightly connected neighborhoods, as cliques and transitive triads, with a higher probability than in random networks \cite{Granovetter1973,Watts1998}. Typically,
friendship networks, or more generally small-world networks, display high cliquishness since two friends of the same person are very likely to be friends\cite{WasFaust, Watts1998}. This behavior made necessary to give a measure for the rate of triangles formation in the network. In a static network, the local clustering coefficient was then quite naturally related to the ratio between actual and potential triangles around a node \cite{Watts1998}.

Many distinct definitions have been proposed for clustering coefficients, particularly for weighted networks, where the choice of the weight attributed to the triangles is crucial \cite{Onnela_2005,Barrat_2004,Opsahl}. Furthermore, recently the identification of the focal node, involved in the definition of clustering coefficient, has been questioned leading to the proposal of a new class of coefficients called \textit{closure coefficients}\cite{Yin2019}.\\ 
While the clustering coefficient quantitatively expresses the idea that two neighbors of the node of interest may be adjacent in turn, the closure coefficient accounts for the fact that the neighbor of a neighbor is, in turn, a neighbor of the focal node. This relation is equally important as the one formalized by the clustering coefficient. However, at the same time, this concept paves the way to a significant increase of the definitions that asks to be systematized in a consistent framework. 

In the present paper, in the more general context of weighted directed multilayer networks, we extend the classical clustering and closure coefficients through the introduction of a new coefficient, here called \textit{clumping coefficient}, which generalizes them to incomplete triangles of any type. Indeed, our argument aims to overcome the rather fictitious and forced distinction between closure and clustering coefficients which is based on a conventional choice in completing open triads of nodes. The first one entails to complete the triangle with the edge opposite to the node of interest, the second one to complete it with the two possible links adjacent to the focal node.\\
Moreover, the introduction of this new coefficient allows us to organize the existing definitions in a systematic taxonomy. In particular, we will show how coefficients in the literature descend as particular cases from a more general definition.

Such coefficients have also been adapted to the different scales that characterizes a multilayer network, in order to grasp their structure from different perspectives, according to the point of view of interest on the network. In particular, local coefficients will be introduced for a single node on a single level, together with coefficients that capture the mesoscale structure of all the replicas of the same node on all levels or of all nodes within a single level. A global coefficient will be also introduced in such a way as to reproduce the classical idea of transitivity for networks.

In the direct case, it is also often useful to have a clear distinction between coefficients that take into account only direct triangles of a certain nature, specifically one of the four types out, in, cycle and middleman triangles. This aspect, although minor and subtle, deserves attention.

To this general purpose, the application of the tensor formalism turns out to be of great help. It allows to reduce all the definitions in the literature and the new ones here introduced to a single unified writing, in such a way that a suitable choice of the involved adjacency tensors allows to obtain each of them.

The papers ends with a thoroughgoing numerical analysis. \added{A detailed example is initially provided in order to highlight the role of the alternative coefficients in the network. Additionally, we have applied} all the indicators previously discussed to random multilayer networks, built according to specific algorithms and with a suitable weights distribution. We show the effectiveness of the new coefficients in capturing different peculiarities of the network structure, at different scales, and we perform a sensitivity analysis \added{to emphasize} their dependence on some key parameters.

The paper is structured as follows: in Section \ref{Mathematics of multilayer networks} we provide the mathematical definitions to manage the multilayer networks using tensors. In Section \ref{Triangles in directed multilayer networks} we deeply describe all types of actual and potential triangles in a multilayer network and we express them in tensors' notation. Section \ref{Clustering for directed multilayer networks} is the core of the work, as we introduce the different types of the new cohesion coefficients, in both binary and weighted cases. In Section  \ref{Results} we test the effectiveness of the provided coefficients by means of the numerical experiments. Conclusions are in Section \ref{Conclusions}. Tables with the formal expression of triangles in terms of tensors, as well as all types of cohesion coefficients proposed in this paper can be found in the Appendix.

\section{Mathematics of directed multilayer networks}
\label{Mathematics of multilayer networks}
\subsection{Basic definitions}

A directed network is formally represented by a graph $G=(V,E)$, where $V$ is a set of $N$ nodes (or vertices) and $E\subseteq V\times V$ a set of arcs (or oriented edges). Two nodes $i$ and $j$ are adjacent if there an arc between $i$ and $j$. If both $(i,j)\in E,$ and $(j,i)\in E$, we say that there is a bilateral arc between $i$ and $j$. Self-loops, i.e. arcs outgoing and incoming in the same node, are not allowed. A directed walk from $i$ to $j$ is a sequence of nodes and arcs with the same direction, starting in $i$ and ending in $j$.  
The binary adjacency relations between pairs of nodes can be conveniently represented by a $N$-square not symmetric matrix $\mathbf{A}$, which is called binary adjacency matrix, whose entries are $a_{ij}=1$ if the ordered pair $(i,j) \in E$, $0$ otherwise. The non-zero entries on the row $i$ of $\mathbf{A}$ represent the arcs coming out of node $i$ (out-matrix). Similarly, the non-zero entries on the row $i$ of $\mathbf{A}^T$ represent the arcs entering in $i$ (in-matrix).  
A labelled directed graph is a graph whose arcs are labelled with symbols. 
The mathematical description of labelled directed graphs is given by $G = (V,E,w)$, where $w$ is a function that assigns a label to each arc. If $w$ assigns non-negative real numbers only, it is called a weight function $w : E \rightarrow {\mathbb R}$ and the corresponding graph is called a weighted directed graph. Adjacency relations are described by the real $N$-square matrix $\mathbf{W}$, the weighted adjacency matrix, of entries $w_{ij}\neq 0$ if there is a weighted arc $(i,j)\in E$, and $w_{ij}= 0$ otherwise.

A weighted \textit{directed multilayer network} (DMN) consists of a family of $L$ weighted directed networks $G_a$, $a=1,...,L$ on the same set of nodes $V$ such that node $i \in G_a$ is said to be adjacent to $j \in G_b$, $\forall a,b=1,...,L$, if there is an oriented edge from $i$ to $j$. Each network $G_a$ is located in a layer in the multilayer network. Arcs can be between nodes in the same layer (intra-layer arcs) or in different layers (inter-layer arcs). When $L=1$, we obtain a single layer network (called {\it monoplex} network).
In particular, we consider a \textit{node-aligned} multilayer network, where all layers share the same set of nodes\footnote{Notice that a non univocal definition of multilayer network is present in the literature, see \cite{Kivela2007} for details}. 

We adopt here the tensor formalism to express the involved mathematical quantities. More specifically, we will reserve Latin letters $i,j,k,..$ and $a,b,c,...$ for objects, namely nodes and layers, respectively, and Greek letters $\mu, \nu, \rho, \sigma, \dots$ and $\alpha, \beta, \gamma, \delta, \dots$ for components of tensors referred to nodes and levels, respectively. $v_{\nu}(i)$ represents the $\nu$ component of a general covariant vector related to node $i$ and $v^{\mu}(i)$ the $\mu$ component of a general contravariant vector related to the same node. 
We define $E^{\mu}_{\nu}(i,j)$ the second order tensor canonical basis in ${\mathbb R}^{N\times N}$, represented by a $N$- square matrix where the $(i,j)$-entry is 1, 0 otherwise. Similarly, $E^{\alpha}_{\beta}(a,b)$ and $E^{\mu \alpha}(i,a)$ are the second order tensor canonical basis in ${\mathbb R}^{L\times L}$ and ${\mathbb R}^{N\times L }$, respectively.

We denote $W^\mu_\nu(a,b)=\sum_{i,j=1}^{N}w_{ij}(a,b)E^{\mu}_{\nu}(i,j)$
the second order inter-layer adjacency tensor for nodes on layers $a$ and $b$, where $w_{ij}(a,b)$ is the weight of the arc from node $i$ on level $a$ to node $j$ on level $b$. When $a=b$ we obtain 
the intra-layer tensor for level $a$, in other words, the weighted adjacency matrix of order $N$ of a monoplex network. The multilayer adjacency tensor is therefore expressed as $M^{\mu \alpha}_{\nu \beta}=\sum_{a,b=1}^{L}W^\mu_\nu(a,b)E^{\alpha}_{\beta}(a,b)$.
Notice that $M^{\mu \alpha}_{\nu \beta}$ is a fourth order tensor encoding all the existing relations between all nodes across all layers. 

Let $A^{\mu \alpha}_{\nu \beta}$ be the binary adjacency tensor, obtained from $M^{\mu \alpha}_{\nu \beta}$ setting all non-zero weights equal to 1. 
The unweighted complete multilayer network is formalized through the adjacency tensor $F^{\mu \alpha}_{\nu \beta}=U^{\mu \alpha}_{\nu \beta}-I^{\mu \alpha}_{\nu \beta}$, where $U^{\mu \alpha}_{\nu \beta}$ is the fourth order tensor whose elements are all equal to 1 and $I^{\mu \alpha}_{\nu \beta}$ is the delta tensor whose elements are equal to 1 if $\mu =\nu$ and $\alpha = \beta$, 0 otherwise. In the complete multilayer network, a node in one level is connected with all its counterparts and all the other nodes in all levels. No self-loops are considered. Then, the unweighted complete directed multilayer network consists of all bilateral arcs with weights equal to $1$. This choice on the weights will be in agreement with the subsequent choice for the normalization of the adjacency tensors.

Throughout the text, for tensors we will adopt the Einstein's summation convention: the summation symbol is omitted for sums over repeated indices. In particular, we will use the tensors contraction, by setting equal a couple of indices, in order to sum with respect to layers, nodes or both. This operation reduces the order of the tensor by 2.

Let us denote by\footnote{Similar conventions will be adopted for the binary tensor $A$}:
\begin{equation}
\begin{split}
& M_{\rm out}=\frac{1}{\mathscr M}M, \quad M_{\rm in}=\frac{1}{\mathscr M}M^T\\
& {\tilde M}=\frac{1}{2}\left( M_{\rm out}+M_{\rm in} \right)
\end{split}
\label{definition}
\end{equation}

and:

\begin{equation}
\begin{split}
&\hat{M}_{\rm out}=\big({M}_{\rm out} \big)^{1/3},\quad \hat{M}_{\rm in}=\big({M}_{\rm in} \big)^{1/3}\\ \quad  &\hat{M}=\frac{1}{2}\left( \hat{M}_{\rm out}+\hat{M}_{\rm in}\right)
	\label{definition2}
\end{split}
\end{equation}

\noindent where $M$ stands for $M^{\mu \alpha}_{\nu \beta}$  ${\mathscr M}=\max_{\mu \nu \alpha \beta}M^{\mu \alpha}_{\nu \beta}$ is the normalisation factor and $M^{T}$ is defined by $\left[{M}^{\mu \alpha}_{\nu \beta}\right]^{T}={M}_{\mu \alpha}^{\nu \beta}$. Note that $M_{\rm out}$ and $M_{\rm in}$ are the natural extension in the tensors context of the out-matrix and in-matrix previously introduced.\\
The definitions provided in formula \eqref{definition2} will allow us to introduce the geometric mean of the arc weights in a triangle.
This will be used later in the paper to extend in the multilayer context the clustering coefficient for directed monoplex networks proposed by \cite{Fagiolo_2007}.

\subsection{Centrality measures: degree and strength}

To introduce the centrality definitions, let us start by observing that, when referring to a weighted DMN, it is appropriate to distinguish an out-strength, an in-strength and a total strength and all of them can be given for a single node $i$ on a specific level $a$ or for all the replicas of node $i$ in all levels. A strength related to the bilateral arcs only can be provided too.

Let us focus, for instance, on the out-strength. We define the out-strength centrality matrix the $N\times L$  matrix whose entries are the out-strengths of each node in each level, $S^{\mu \alpha}_{\rm out}=M^{\mu \alpha}_{\nu \beta}u^{\beta}u^{\nu}$, being $u^{\beta}$ and $u^{\nu}$ the $L$-vector and the $N$-vector of all $1$'s, respectively. The global out-strength of node $i$ on the whole multilayer network, taking into account all nodes $i$ on all the $L$ levels is expressed by the vector $s^{\mu}_{\rm out}=M^{\mu \alpha}_{\nu \beta}U^{\beta}_{\alpha}u^{\nu}$
, where $U^{\beta}_{\alpha}$ is the $L$-square matrix of all $1$'s. 
Similarly, for the in-strengths, we have: $S_{\nu \beta}^{\rm in}=u_{\alpha}u_{\mu}M^{\mu \alpha}_{\nu \beta}$ and $s_{\nu}^{\rm in}=u_{\mu}M^{\mu \alpha}_{\nu \beta}U^{\beta}_{\alpha}$. 
The total strength $S_{\rm tot}$ is the sum of the out- and the in-strength.
Analogous definitions can be provided for in- and out-degree. We denote by $K^{\mu \alpha}_{\rm out}$, $k^{\mu}_{\rm out}$ and so on the degree centrality measures.

\section{Triangles in directed multilayer networks}
\label{Triangles in directed multilayer networks}

Generally speaking, a triangle in a multilayer network is a closed triplet $i,j,k$ such that the three nodes can belong to up to three different layers and they are connected by inter or intra-layer arcs, independently of their orientation. 
If no cost is associated with the jump from one level to another, in a multilayer it can be considered triangles of general type, where the three arcs could lie on different layers, being connected by up to three further arcs between levels.\\
However, we assume here a definition that extends the proposal of \cite{DeDomenico2013} for a monoplex unweighted network. It includes all possible closed triplets, moving in all directions, along inter or intra-layer links, but composed by exactly three arcs. In direct multilayer networks, as well as in monoplex networks, we can identify four types of triangles around a given node $i$ on layer $a$ according to the orientation of the arcs. We refer to the classification provided in the literature (see e.g. \cite{Fagiolo_2007}), namely out, in, cycle and middleman triangles.

\subsection{Actual triangles}
The number of actual triangles of each type can be calculated by counting the accordingly oriented $3$-cycles around the node of interest. 
The total number $T_{\rm tot}^{\rm act}(i,a)$ of actual triangles around node $i$ on layer $a$ is defined as:

\begin{equation}
T_{\rm tot}^{\rm act}(i,a) =
4{\tilde A}^{\mu \alpha}_{\nu \beta}{\tilde A}^{\nu \beta}_{\rho \gamma}{\tilde A}^{\rho \gamma}_{\sigma \delta}E_{\mu \alpha}(i,a)E^{\sigma \delta}(i,a)
\end{equation}

\noindent being the binary tensor $\tilde A$ defined as in formula (\ref{definition}). Note that the number $4$ in the formula is aimed at compensating for the factor $1/2$ in the definition of  $\tilde A$.
Table \ref{TableA1} (see the Appendix) provides the number of all actual triangles in a specific class, referred to a single node on a single level.

However, when dealing with 
a DMN, we could be interested in the number of triangles to which all the replicas of given node belong, or in all triangles formed by all nodes within a level. In this perspective, the representation by tensors encodes a flexibility that allows
to grasp different observation scales on the network, depending on suitable contractions on the tensors involved. 
Indeed, by contracting over all the levels on which a node $i$ lies, we obtain the total number $T_{\rm tot}^{\rm act}(i)$ of triangles to which $i$ belongs:

\begin{equation}\label{triangles_node}
	T_{\rm tot}^{\rm act}(i)=4{\tilde A}^{\mu \alpha}_{\nu \beta}{\tilde A}^{\nu \beta}_{\rho \gamma}{\tilde A}^{\rho \gamma}_{\sigma \alpha}E_{\mu}^{\sigma}(i,i).
\end{equation}

By contracting over all the nodes on the same layer $a$ we obtain the number $T_{\rm tot}^{\rm act}(a)$ of triangles to which all nodes on level $a$ belong

\begin{equation}\label{triangles_level}
	T_{\rm tot}^{\rm act}(a)=4{\tilde A}^{\mu \alpha}_{\nu \beta}{\tilde A}^{\nu \beta}_{\rho \gamma}{\tilde A}^{\rho \gamma}_{\mu \delta}E_{\alpha}^{\delta}(a,a)
\end{equation}

Finally, by contracting over all nodes and layers, we obtain the total number of triangles:

\begin{equation}\label{triangles}
	T_{\rm tot}^{\rm act}=4{\tilde A}^{\mu \alpha}_{\nu \beta}{\tilde A}^{\nu \beta}_{\rho \gamma}{\tilde A}^{\rho \gamma}_{\mu \alpha}
\end{equation}

\begin{figure*}
	\centering
	\includegraphics[scale=0.35]{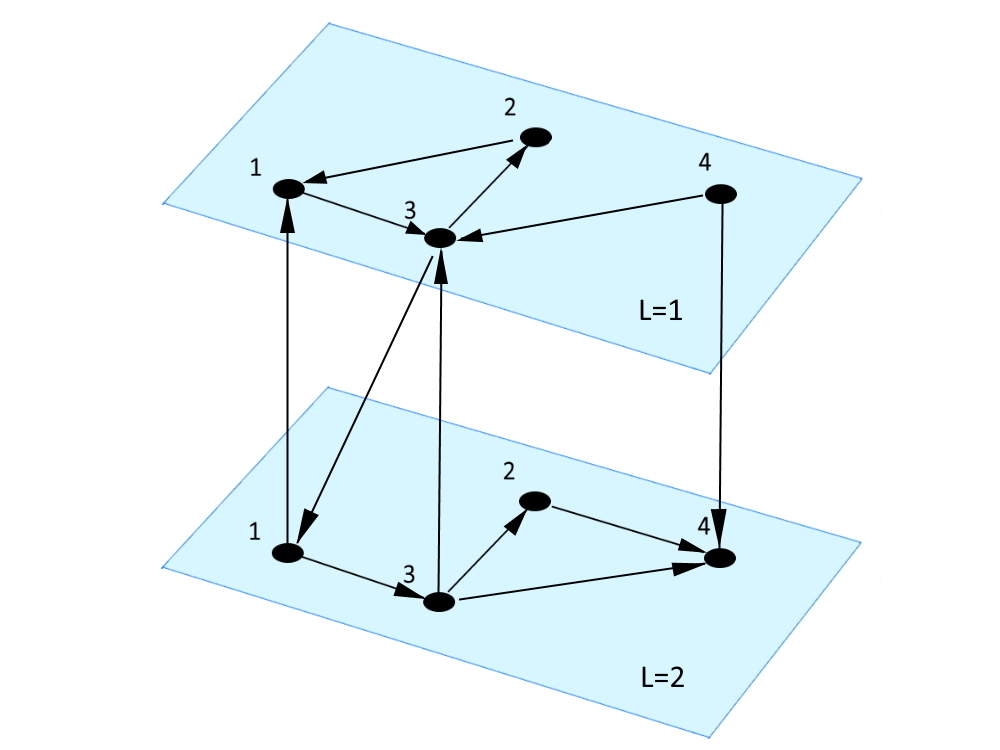}
	\caption{Example of a binary DMN with 4 nodes and 2 layers}
	\label{fig2}
\end{figure*}

The types of triangles existing in a multilayer are represented by the simple network in Figure \ref{fig2}, where four nodes on two distinct levels are connected by oriented arcs. For instance, node $1$ on both layers 1 and 2 belongs to 2 cycles. Node $2$ belongs to a cycle on layer 1 and to a middleman on layer 2.
 
\subsection{Potential triangles}
\begin{figure*}
	\centering
	\includegraphics[scale=0.5]{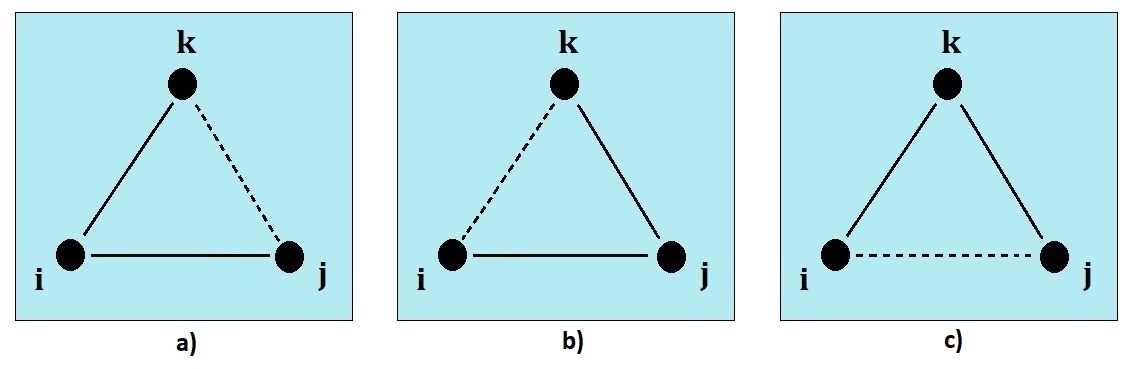}
	\caption{The three alternative ways to complete an open triad, where node $i$ is the focal node 
	}
	\label{fig3}
\end{figure*}

The question immediately arises of how to complete an open triplet, generating a triangle. The first possibility is that nodes sharing a common neighbor $i$ are in turn connected: in this case the open triad is closed by the link opposite to $i$ (see Figure \ref{fig3} (a), where node $i$ is the central node). The second possibility occurs when the open triad is closed by connecting $i$ with the neighbor of a neighbor of  $i$ (see Figure \ref{fig3} (b) and (c), where $i$ is the peripheral node).

\begin{figure*}
	\centering
	\includegraphics[scale=0.4]{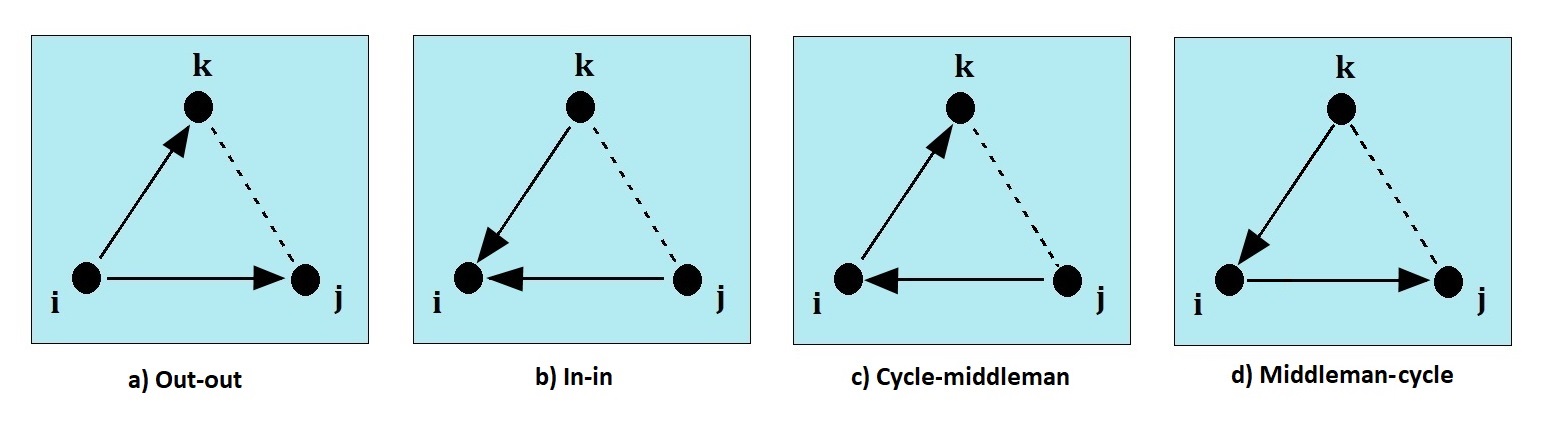}
	\caption{Alternative open directed triads, where node $i$ is the focal node}
	\label{fig4}
\end{figure*}

In a weighted 
DMN a weight can be assigned to a triangle according to a non univocal choice. 
Since the weighted tensors we have introduced are all normalized with the maximum weight, it is reasonable to complete a triad with a bilateral arc of weight equal to $1$. With this choice, we include both possible orientations and attribute the maximum weight to the missing link. In Figure \ref{fig4} we illustrate all the potential triangles where node $i$ is the central one. In particular, in panels (a) and (b) two potential out-triangles and in-triangles can be obtained completing the link between $k$ and $j$ with two arcs of weight 1. In each panel (c) and (d) we can obtain a cycle and a middleman triangle completing the missing link with two arcs of weight 1.  Notice that the number of all possible triangles depicted in Figure \ref{fig4} can be computed as product of three appropriate adjacency tensors. The completion of the missing link can be formalized by using the tensor $F^{\mu \alpha}_{\nu \beta}$. The number of triangles of each class is reported in Table \ref{TableA3}, in the Appendix.\\

The total number of potential triangles of this first type is computed as: 
\begin{equation}
T_{\rm tot}^I(i,a)=
4{\tilde A}^{\mu \alpha}_{\nu \beta}{F}^{\nu \beta}_{\rho \gamma}{\tilde A}^{\rho \gamma}_{\sigma \delta}E_{\mu \alpha}(i,a)E^{\sigma \delta}(i,a).
\label{pot_triangles_I}
\end{equation} 

As shown in Table \ref{TableA3}, the total number $T_{\rm tot}^I(i,a)$ can be obtained as the sum of the total number of triangles of different classes.  Note that the number of potential cycles is equal to the number of potential middleman triangles. 

\begin{figure*}
	\centering
	\includegraphics[scale=0.4]{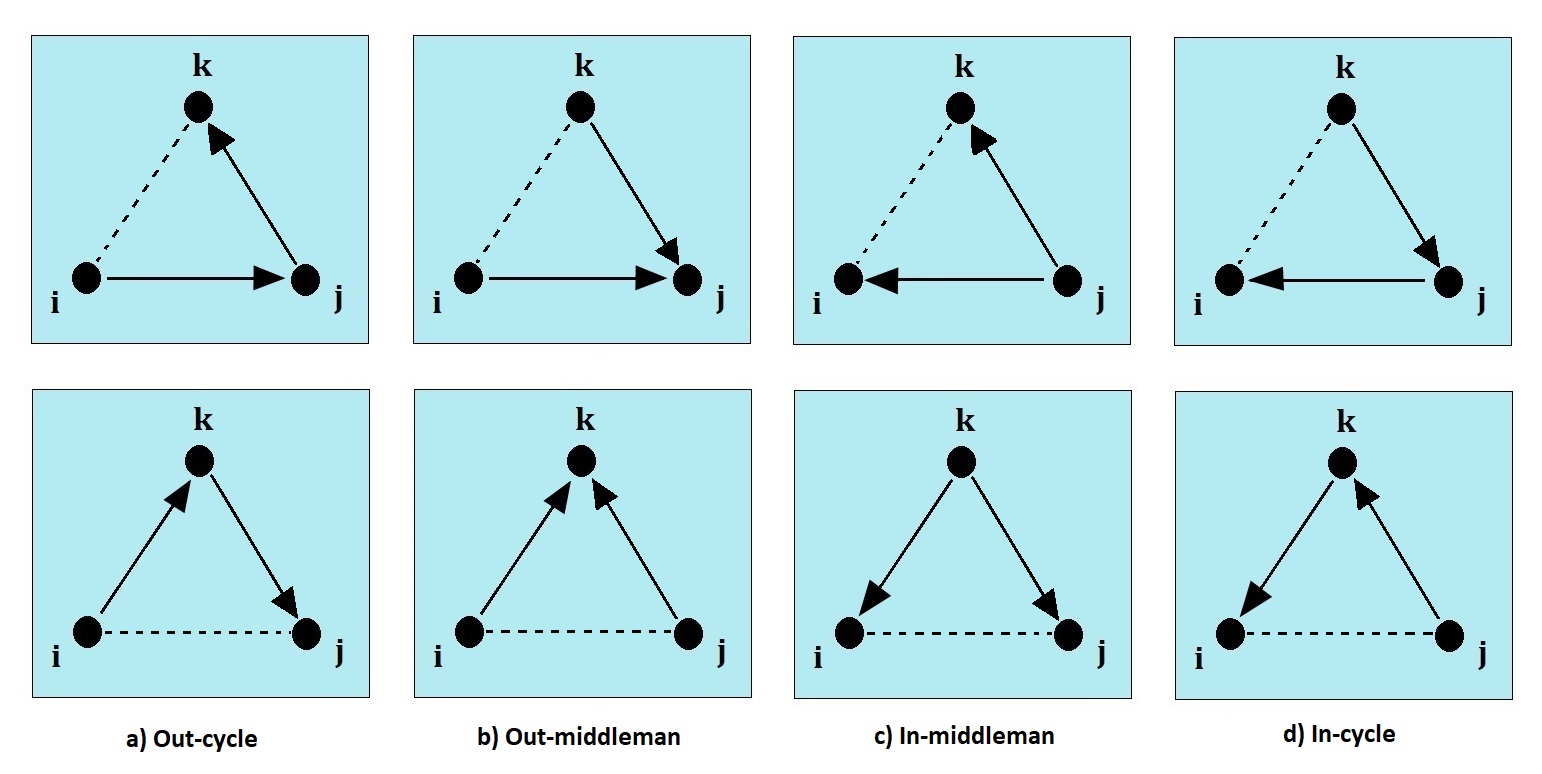}
	\caption{Alternative ways to complete open directed triads, when the focal node $i$ is peripheral. Each panel refers to the two graphs above the caption}
	\label{fig5}
\end{figure*}

Let us focus on the second type of potential triangles, where the focal node $i$ is peripheral, i.e. one of the two possible ends of the open triad. Figure \ref{fig5}  illustrates all possible cases. Note that the presence of the potential bilateral arc generates two different types of directed triangles, depending on the direction in which this link is crossed. For instance, in panel (a), according to the direction of the missing link,  we can obtain four triangles, namely two cycles and two out-triangles.

The total number of potential triangles of this second type is: 

\begin{equation}
T_{\rm tot}^{II}(i,a)=
4{F}^{\mu \alpha}_{\nu \beta}{\tilde{A}}^{\nu \beta}_{\rho \gamma}{\tilde A}^{\rho \gamma}_{\sigma \delta}E_{\mu \alpha}(i,a)E^{\sigma \delta}(i,a)
+4{\tilde{A}}^{\mu \alpha}_{\nu \beta}{\tilde{A}}^{\nu \beta}_{\rho \gamma}{F}^{\rho \gamma}_{\sigma \delta}E_{\mu \alpha}(i,a)E^{\sigma \delta}(i,a).
\label{pot_triangles_II}
\end{equation}

Also in this case, the total number $T_{\rm tot}^{II}(i,a)$ is equal to the sum of the number of triangles of different classes (see Table \ref{TableA3} in the Appendix). For both types of triangles (i.e. I and II type) it is possible to adapt formulas (\ref{pot_triangles_I}) and (\ref{pot_triangles_II}) in order to compute the total number of potential triangles for the node $i$, the layer $a$ or for the whole network.

\section{Clustering, closure and clumping coefficients in DMNs}
\label{Clustering for directed multilayer networks}

\subsection{Clustering Coefficients for binary DMNs}
\label{General definitions}

Local clustering coefficients are in general defined as the ratio between the number of actual triangles to which node $i$ belongs and the number of potential triangles $i$ could form with its neighbors, that is the number of open triads in which $i$ is the central node \cite{Watts1998}. 

We define the \textit{clustering coefficients} as:
\begin{equation}\label{e:clust_nodo_livello}
C^{\rm clust}(*)=\frac{T_{\rm tot}^{act}(*)}{T_{\rm tot}^I(*)}
\end{equation}

\noindent where $\left\{*\right\}=\left\{ (i,a),(i),(a),(.) \right\}$, being $(.)$ the notation for the global coefficient.

Notice that formula (\ref{e:clust_nodo_livello}) encodes different coefficients in the unweighted case. In particular, varying the number of actual and potential triangles considered, we can obtain the clustering coefficient for a node $i$ in a level $a$ $C^{clust}(i,a)$, for a node $i$ on all the levels $C^{clust}(i)$, for all the nodes in the level $a$ $C^{clust}(a)$, and the global coefficient in the network $C^{clust}$.

Considering instead the second type of potential triangles, we can focus on a local closure coefficient. Local closure coefficients are defined in the literature as the ratio between twice the number of actual triangles containing $i$ and the number of open triads that end in $i$ (\cite{Yin2019, Jia2021}). \\
We define the \textit{closure coefficients} as:
\begin{equation}\label{e:clos_nodo_livello}
C^{\rm clos}(*)=\frac{2T_{\rm tot}^{act}(*)}{T_{\rm tot}^{II}(*)}
\end{equation}
\noindent where $\left\{*\right\}=\left\{ (i,a),(i),(a),(.) \right\}$.
Also in this case, we provide a general definition that allows to consider triangles under different perspectives (node, levels, etc.). \\
Hence, clustering coefficients return the percentage of real triangles to the number of open triads centered in $i$ (as shown before in Figure \ref{fig3}, panel a), whereas closure coefficients return the percentage of real triangles to the number of open triads ending in $i$ (see Figure \ref{fig3} panels b and c). The number of actual triangles, in this case, is counted twice, since each triangle contains two triads that end in $i$ (i.e. $ijk$ and $jki$).

To provide a complete view, we propose a third coefficient called \textit{clumping coefficient}, that takes into account all types of triangle present in the network. It is defined as follows:
\begin{equation}\label{e:clump_nodo_livello}
C^{\rm clump}(*)=\frac{3T_{\rm tot}^{act}(*)}{T_{\rm tot}^I(*)+T_{\rm tot}^{II}(*)}
\end{equation}
\noindent where $\left\{*\right\}=\left\{ (i,a),(i),(a),(.) \right\}$.

It is worth noting that the clumping coefficient can be obtained as a weighted average of the clustering and closure coefficients. 
Indeed, formula (\ref{e:clump_nodo_livello}) can be written as

\begin{equation}
C^{\rm clump}(*)=
\frac{T_{\rm tot}^{I}(*)C^{\rm clust}(*)+T_{\rm tot}^{II}(*)C^{\rm clos}(*)}{T_{\rm tot}^I(*)+T_{\rm tot}^{II}(*)}
\label{e:clump_nodo_livello_mediap}
\end{equation}

It is noteworthy that formulas (\ref{e:clust_nodo_livello}-\ref{e:clump_nodo_livello}) consider all the triangles in the network. The same formulas can be computed considering only a specific class. For instance, the out-clustering coefficient $C^{clust}_{out}(i,a)$ for a node $i$ at level $a$ can be computed considering the ratio between actual out-triangles and potential out-triangles of the first type for the node $i$ in the level $a$ (see Tables \ref{TableA1} and \ref{TableA3} in Appendix). In a similar way, we can obtain coefficients for other classes (in, cycle and middleman),  aggregated at different levels (node on all levels, all node in a level, global) or for different types (clustering, closure, clumping).

\subsection{Clustering coefficients for weighted DMNs}
\label{Clustering coefficients for weighted DMN}

The coefficients previously defined for binary DMNs can be extended to weighted DMNs, once we decided on which way to assign the weighs to both the actual and potential triangles.
For the sake of brevity, in this section we refer to the clustering coefficient $C^{clust}(i,a)$ for a node $i$ at level $a$, but similar results hold for the other cases and for both closure and clumping coefficients. 

Formally, the weighted clustering coefficient is still defined as the ratio between the actual and the potential triangles, like in formula (\ref{e:clust_nodo_livello}). 
However, considering the weighted version, the general definition generates different coefficients, depending on the way in which we assign the weights to the involved triangles.

Both the actual and potential triangles can be weighted using the product of the weights. Formally, the number of actual triangles $T_{\rm tot}^{\rm act}=4{\tilde A}{\tilde A}{\tilde A}$ (as reported in the Appendix, Table \ref{TableA1})  is replaced by the tensors product ${4\tilde M}{\tilde M}{\tilde M}$. The number of all potential triangles $T_{\rm tot}^{\rm I}=4{\tilde A}F{\tilde A}$ (as in Table \ref{TableA3}) is replaced by the tensors product $4\tilde{M} F \tilde{M}$.
We then define the weighted clustering coefficient $\tilde{C}^{\rm clust}(i,a)$ as:

\begin{equation}\label{DeDomenico}
\tilde{C}^{\rm clust}(i,a)=\frac{\tilde{M} \tilde{M} \tilde{M}}{\tilde{M} F \tilde{M}}
\end{equation}
	
Observe that, in formula (\ref{DeDomenico}), the weight attributed to each triangle is obtained as the product of the weights of the three arcs, namely by the product of the corresponding components in the normalized tensor $M$. Note that formula (\ref{DeDomenico}) extends to a DMN the clustering coefficient introduced in \cite{DeDomenico2013} for the global case in an undirected multilayer network.

Alternatively, both the actual and potential triangles can be weighted using the arithmetic mean of the weights of the arcs incident to the node $i$. In this case, $T_{\rm tot}^{\rm act}$ and $T_{\rm tot}^{\rm I}$ are equal to $\tilde{M} \tilde{A} \tilde{A}$ and ${\tilde{M} F \tilde{A}}$, respectively. We then define the weighted clustering coefficient $C^{\rm clust}(i,a)$ as:
		
\begin{equation}\label{ClementeGrassi}
	C^{\rm clust}(i,a)=\frac{\tilde{M} \tilde{A} \tilde{A}}{\tilde{M} F \tilde{A}}
\end{equation}
	
As already pointed out, the weight attributed to each triangle is the mean of the weights of the arcs between node $i$ and its neighbors. In other words, it considers only two of the three links involved in the closed triplet, namely those adjacent to node $i$. Indeed, the tensor product in the numerator considers these two weights and it requires the existence of the third link between the neighbors of the node $i$.	Hence, formula (\ref{ClementeGrassi}) extends to the multilayer case the clustering coefficient proposed in \cite{CleGra} for monoplex directed networks.\footnote{In a monoplex network the local clustering coefficient defined by  \cite{CleGra} is:
$$
c(i)=\frac{\big[ \left( \mathbf{W}+\mathbf{W}^{T} \right)\left(\mathbf{A}+\mathbf{A}^{T} \right)^{2}\big]_{ii}}{2\left[ s_{i}^{\rm tot}\big(k_{i}^{\rm tot}-1\big)-2s_{i}^{\rm bil}\right] }
$$ where $k_{i}^{\rm tot}$, $s_{i}^{\rm tot}$, $s_{i}^{\rm bil}$ are the total degree, strength and bilateral strength of node $i$, respectively.}

Finally, only the actual triangles can be weighted using the geometric mean of the weights. 
In this case, $T_{\rm tot}^{\rm act}$ and $T_{\rm tot}^{\rm I}$ are equal to $\hat{M} \hat{M} \hat{M}$ and $\hat{A} F \hat{A}$, respectively. We then define the weighted clustering coefficient $\hat{C}^{\rm clust}(i,a)$ as:
 	
\begin{equation}\label{Fagiolo}
	\hat{C}^{\rm clust}(i,a)=\frac{\hat{M} \hat{M} \hat{M}}{\hat{A} F \hat{A}}
\end{equation}

Note that in formula (\ref{Fagiolo}) the numerator contains the geometric mean of the corresponding components of the normalized tensor $M$. The denominator counts the number of potential triangles to which $i$ belongs, ignoring the weights. 
This fact makes this coefficient the immediate generalization to directed multilayer networks of the coefficient proposed in \cite{Fagiolo_2007} for monoplex networks.\footnote{In a monoplex network the local clustering coefficient defined by \cite{Fagiolo_2007} is given by:
$$
\hat{c}(i)=\frac{\left( \mathbf{\tilde W}+\mathbf{\tilde W}^{T} \right)^3_{ii}}{2\left[ k_{i}^{\rm tot}\big(k_{i}^{\rm tot}-1\big)-2k_{i}^{\rm bil}\right] }
$$ where $\tilde {\mathbf{W}} = [\frac{\mathbf{W}}{\max_{ij}(w_{ij}) }]^{\frac{1}{3}}$ and $k_{i}^{\rm bil}$ is the total bilateral degree of node $i$.}

We collect in the Appendix the explicit expressions for all the versions of clustering, closure and clumping coefficients in weighted DMNs for a node $i$ on level $a$.

\section{Results}
\label{Results}
We provide in this section \added{a detailed example to illustrate the different role played by the alternative coefficients proposed and their interpretation in a social media context.} 
\added{Additionally, a numerical analysis is also developed} to test the behavior of the coefficients \added{on different classes of graphs. In particular, we consider} a random graph, based on Erdos-Renyi (ER) model (see \cite{ER1959} and \cite{ER1960}), and a small-world (SW) network, based on Watts and Strogatz (see \cite{Watts1998}) model. \added{S}ome sensitivity analyses have been \added{also} exploited by testing the effect of some key parameters (as the number of nodes, layers, rewiring probabilities, etc.).

\subsection{\added{A numerical example}}

\begin{figure}[H]
	\centering
	\includegraphics[scale=0.3]{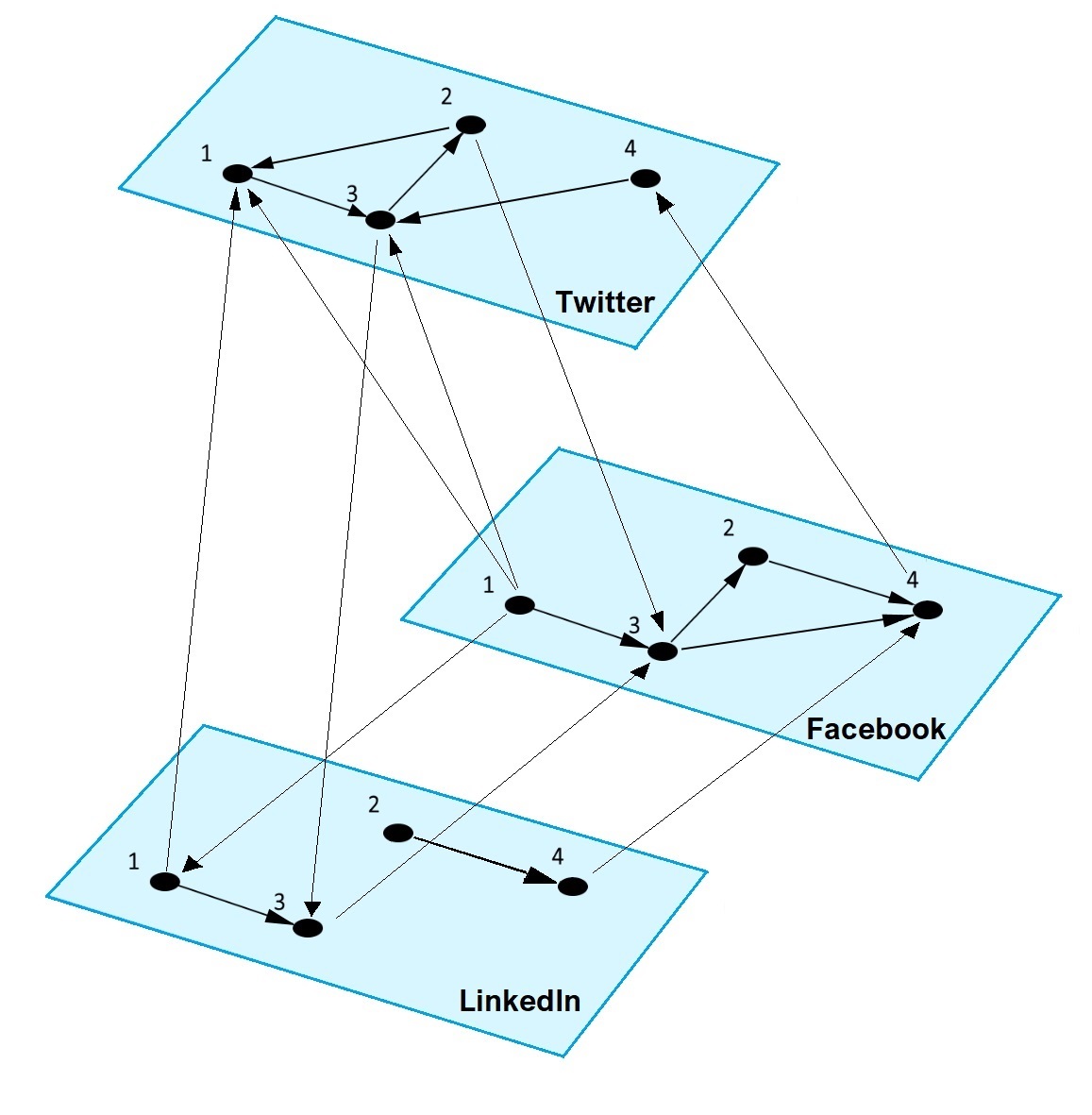}
	\caption{Example of a Binary Directed Multilayer Network with 4 nodes and 3 layers}
	\label{fig6}
\end{figure}

\added{We consider a multilayer formulation of the mention or tag network on Twitter, Facebook and LinkedIn. A mention is the inclusion within a tweet or a post of a name containing a hyperlink to a different subject, typically another account's Twitter user name, preceded by the @ symbol, or another individual profile on Facebook and LinkedIn. Twitter, Facebook and LinkedIn represent the three layers. We consider in each layer the same set of subjects or profiles, assuming that they all have accounts on all three social media. When a subject publishes a tag within a social media, i.e. a name containing a hyperlink within the same social media or to one of the other two, an arc is created from the node/subject within the publishing media to the node/subject in the media to which the hyperlink refers. The arc is oriented from the publishing media to the target media. For instance, if subject A posts on Facebook a message containing "@subjectB" toward Twitter, an arc is create from node A on layer Facebook to node B on layer Twitter. Of course, subject A can post on Facebook a message containing "@subjectA" toward Twitter, creating an arc between aligned nodes on different layers. If we examine a given time interval, we can assign a weight to the arcs: the weight is given by the number of instances when subject A in the layer $\alpha$ generates a tag toward subject B in the layer $\beta$. Therefore, we are dealing with a proper directed multilayer network.
Let us consider, for instance, four subjects, identified by nodes $1$, $2$, $3$ and $4$, that are all present on the three different social media (i.e. the layers). 
We depict the structure of the tag network in Figure \ref{fig6}.
For the sake of simplicity, we refer here only to the binary version of the network, deferring to subsequent numerical analysis the impact of the weights on the values of the different coefficients.}

\added{We collect in Table \ref{Table1} the number of triangles of each type for a single node on a single level, obtained by relations in Table \ref{TableA1}, and the total number of actual triangles for each single subject and each single layer in the whole multilayer network.}

\begin{table}[H]
	\scriptsize
	\centering{}
	\renewcommand{\arraystretch}{1.2}
	\begin{tabular}{|c||c|c|c|c||c|c|c|c||c|c|c|c||c|c|c|c||}
		\hline \hline
		\multicolumn{17}{|c|}{\bf Number of triangles}\tabularnewline \hline 
		{\bf Layer} & \multicolumn{4}{|c||}{\bf Twitter} & \multicolumn{4}{|c||}{\bf Facebook} & \multicolumn{4}{|c||}{\bf LinkedIn} & \multicolumn{4}{|c||}{\bf Network} \tabularnewline \hline
		{\bf Node}  & {\bf 1} & {\bf 2} & {\bf 3} & {\bf 4} & {\bf 1} & {\bf 2} & {\bf 3} & {\bf 4} & {\bf 1} & {\bf 2} & {\bf 3} & {\bf 4} & {\bf 1} & {\bf 2} & {\bf 3} & {\bf 4}  \tabularnewline \hline
		$T_{\rm out}^{\rm act}$  & 0 & 0 & 0 & 0 & 2 & 0 & 1 & 0 & 0 & 0 & 0 & 0 & \multirow{5}{*}{6} & \multirow{5}{*}{2} & \multirow{5}{*}{3} & \multirow{5}{*}{1} \tabularnewline \cline{1-13}
		$T_{\rm in}^{\rm act}$ & 1 & 0 & 1 & 0 & 0 & 0 & 0 & 1 & 0 & 0 & 0 & 0 & & & &  \tabularnewline \cline{1-13}
		$T_{\rm cycle}^{\rm act}$  & 1  & 1 & 1 & 0 & 0 & 0 & 0 & 0 & 0 & 0 & 0 & 0 & & & &  \tabularnewline \cline{1-13}
		$T_{\rm mid}^{\rm act}$  & 1 & 0 & 0 & 0 & 0 & 1 & 0 & 0  & 1 & 0 & 0 & 0 & & & & \tabularnewline \cline{1-13}
		$T_{\rm tot}^{\rm act}$  & 3 & 1 & 2 & 0 & 2 & 1 & 1 & 1 & 1 & 0 & 0 & 0 & & & & \tabularnewline \hline
		${\rm Total}$ & \multicolumn{4}{|c||}{6} & \multicolumn{4}{|c||}{5} & \multicolumn{4}{|c||}{1} & \multicolumn{4}{|c||}{12} \tabularnewline \hline
		\hline
	\end{tabular}\caption{Number of actual triangles in the network illustrated in the text.}
	\label{Table1}
\end{table}

\added{In Tables \ref{Table2} and \ref{Table3}, we collect the values of the local clustering, closure and clumping coefficients for each node in each level as in Eqs. \ref{e:clust_nodo_livello}, \ref{e:clos_nodo_livello} and \ref{e:clump_nodo_livello}. Let us remind that, for a binary network, $\tilde{C}^{\rm clust}(*)={C}^{\rm clust}(*)=\hat{C}^{\rm clust}(*)$, and similarly for closure and clumping coefficients.}

\begin{table}[H]
	\scriptsize
	\centering{}
	\begin{tabular}{|c||c|c|c|c||c|c|c|c||c|c|c|c||}
		\hline \hline
		\multicolumn{13}{|c|}{\bf Local Coefficients}\tabularnewline
		\hline 
		\multirow[c]{2}{*}{\bf Version} & \multicolumn{4}{c||}{\bf Twitter} & \multicolumn{4}{c||}{\bf Facebook} & \multicolumn{4}{c||}{\bf LinkedIn} \tabularnewline
		\cline{2-13}
		& 1 & 2 & 3 & 4  & 1 & 2 & 3 & 4 & 1 & 2 & 3 & 4 \tabularnewline
		\hline 
		\centering
		{\bf Clustering} & $0.250$ & $0.167$ & $0.100$ & $0.000$ & $0.167$ & $0.500$ & $0.050$ & $0.083$ & $0.167$ & $-$ & $0.000$ & $0.000$  \tabularnewline
		{\bf Closure}  & $0.273$ & $0.091$ & $0.182$ & $0.000$ & $0.154$ & $0.143$ & $0.091$ & $0.143$ & $0.125$ & $0.000$ & $0.000$ & $0.000$ 
		\tabularnewline
		{\bf Clumping}  & $0.265$ & $0.107$ & $0.143$ & $0.000$ & $0.158$ & $0.188$ & $0.071$ & $0.115$ & $0.136$ & $0.000$ & $0.000$ & $0.000$ 
		\tabularnewline
		\hline \hline
	\end{tabular}
	\caption{Local Coefficients for each node in each level}
	\label{Table2}
\end{table}

\begin{table}[H]
	\scriptsize
	\centering{}
	\begin{tabular}{|c||c|c|c|c||c|c|c||c|}
		\hline \hline
		\multicolumn{9}{|c|}{\bf Node, Levels and Global Coefficients}\tabularnewline
		\hline 
		\multirow[c]{2}{*}{\bf Version} & \multicolumn{4}{c||}{\bf Nodes} & \multicolumn{3}{c||}{\bf Levels} & \multirow[t]{2}{*}{\bf Whole}\tabularnewline
		\cline{2-8}
		& {\bf 1} & {\bf 2} & {\bf 3} & {\bf 4} & {\bf Twitter} & {\bf Facebook} & {\bf LinkedIn} & {\bf Network}\tabularnewline
		\hline 
		\centering
		{\bf Clustering} & $0.200$ & $0.250$ & $0.065$ & $0.063$ & $0.150$ & $0.109$ & $0.071$ & $0.120$ \tabularnewline
		{\bf Closure}  & $0.188$ & $0.105$ & $0.094$ & $0.059$ & $0.150$  & $0.131$ & $0.045$ & $0.120$ \tabularnewline
		{\bf Clumping} & $0.191$ & $0.130$ & $0.082$ & $0.060$ & $0.150$ & $0.123$ & $0.052$ & $0.120$ \tabularnewline
		
		\hline \hline
	\end{tabular}
	\caption{Coefficients for each node $i$, for each level $a$ and for the whole network}
	\label{Table3}
\end{table}

\added{In order to make the numerical results in Tables \ref{Table2} and \ref{Table3} meaningful, 
we offer the following interpretation of them.\\
The clustering coefficient returns the probability that, if a subject tags two others, they in turn will be induced to tag each other. The focal node, which is central, acts actively and induces the triangle completion between the two contacts.
The closure coefficient returns the probability that, if a tag links to a post containing a second tag, then the first subject is induced in turn to tag this third one. The focal node, which is in this case one of the ends in triads, acts passively and the network induces or pushes it to close the triangle.\\
Thus, the two coefficients emphasize two complementary aspects in the behavior of the focal subject. 
A high value of the clustering coefficient means that a large fraction of the potential triangles are closed by the two neighbors of the focal central node. So, such a node wields a strong influence on its neighbors and induces them to create links between them. This implies an \textit{influencer }action of the node on its contacts.\\
If, conversely, a node has a high closure coefficient then it tends to close triangles with high probability: if a friend creates a link with a third one, it tends to be influenced thereby closing the link with this third one. Thus such a node is characterized as being more likely to be influenced.\\
On the one hands, if the focal node belongs to a great number of potential triangles as a central node and few of them are closed by neighbors, the clustering coefficient is low and this subject is not effective in influencing its neighbors to create a new link. In other words, it is a bad influencer.
On the other hands, if the focal node belongs to a great number of potential triangles as end node and it closes a few of them with its second-order neighbors, the closure coefficient is low and this subject does not tend to be influenced by the links between his neighbors and others and thus appears more difficult to influence.\\
Finally, the clumping coefficient returns the probability that a subject is a part of a triad, either because he induced neighbors to link each others or because others induced him to generate a new link. 
Focusing on Twitter and observing Tables \ref{Table2} and \ref{Table3} for the  network instance in figure \ref{fig6}, we see that subject $1$ exerts a strong influencer action on the other subjects inside and outside the layer but, at the same time, it is also strongly influenced by them, as evidenced by the high value of clumping coefficient. In contrast, nodes $2$ and $3$ appear to have complementary behaviors: subject $2$ acts as influencer more than being influenced, while subject $3$ appears more influenced by the others.
Considering now the behavior of the subjects in all the social media, subject $2$ is the main influencer because of the highest clustering coefficient.
However the most active node in completing triads in all possible ways remains node $1$, given the highest value of the clumping coefficient. Considering instead the role of the alternative social media, Twitter turns out to be the most influential platform, with equal contribution in terms of closing central or peripheral triads, while LinkedIn appears to be the least active in the global multilayer network.}\footnote{It should be observed that a pendant node, like node $2$ on the LinkedIn layer, does not belong to any real or potential triangle as a central node and therefore its clustering coefficient cannot be calculated. Sometimes a conventional value of zero is assigned to it. When a node, although belonging to one or more potential triangles, does not belong to any real triangle, as a central or end node, the corresponding coefficient is zero. This is the case of node $4$ on the layer Twitter or node $2$ on the layer Facebook. However, it is appropriate to clearly distinguish the two cases.}

\subsection{\added{Erdos-Renyi graphs}}
\begin{figure*}
	\centering
	\includegraphics[width=12cm,height=8cm]{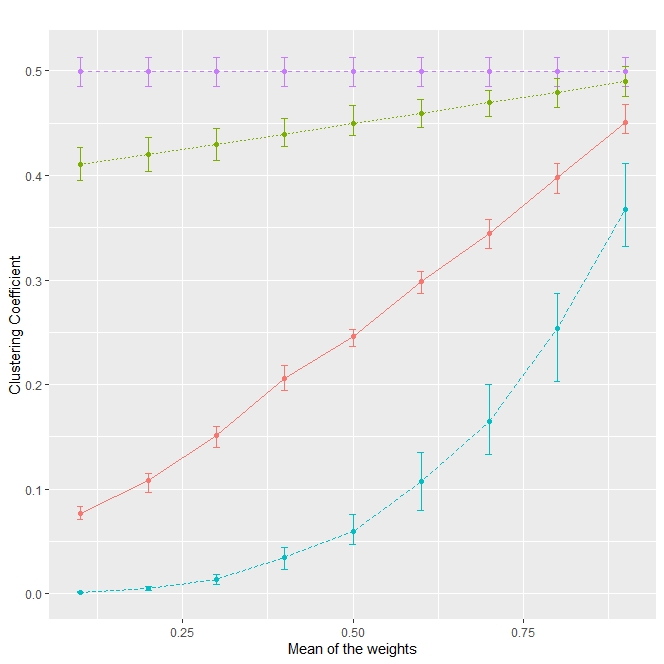}
	\caption{Comparison of clustering coefficients for a ER multilayer network, 50 nodes, 2 layers and attachment probability equal to 0.5. In green, red and light blue we report the distributions of $C(i,\alpha)$, $\tilde{C}(i,\alpha)$  and $\hat{C}(i,\alpha)$ for each choice of the weight. For each combination, vertical bars denote the range of the values, while the dot is the global coefficient for the whole network. Purple behavior is related to the coefficients computed on the unweighted network. In this case the three coefficients give the same results and do not depend on weights.
	}
	\label{ClustER}
\end{figure*}

\begin{figure*}
	\centering
	\includegraphics[width=6cm,height=6cm]{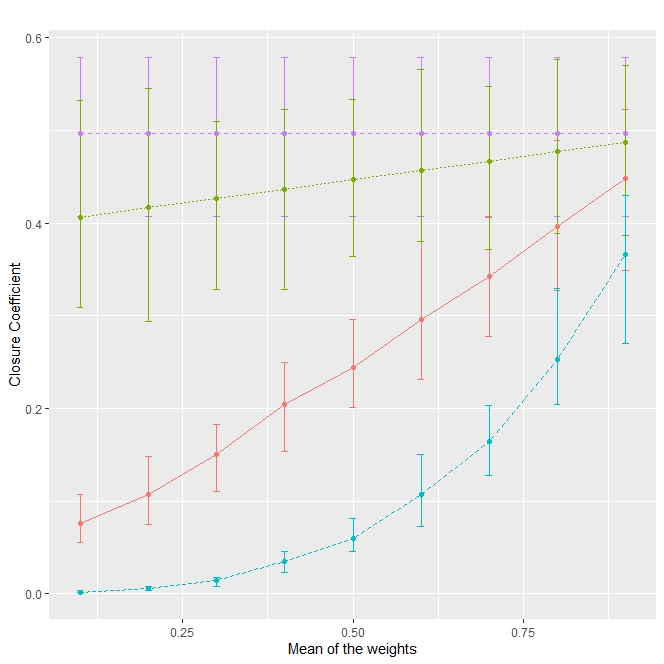}
	\includegraphics[width=6cm,height=6cm]{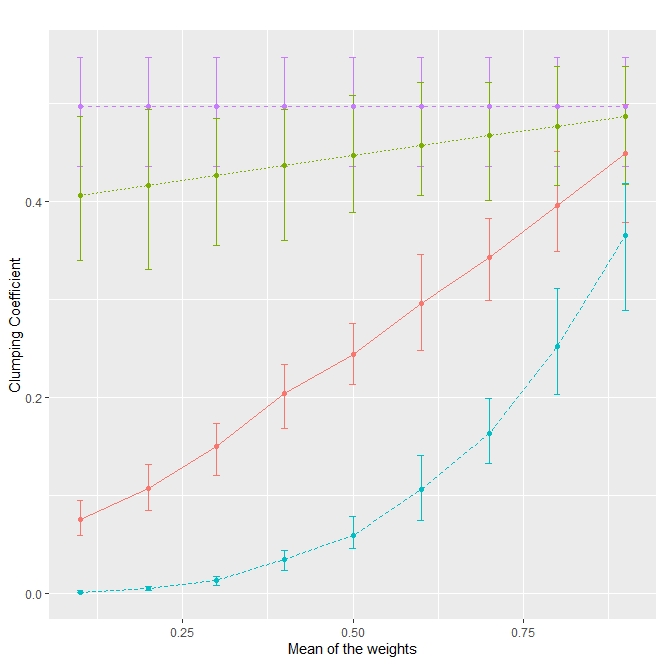}
	\caption{Comparison of closure and clumping coefficients for a ER multilayer network, 50 nodes, 2 layers and attachment probability equal to 0.5. In green, red and light blue we report the distributions of $C(i,\alpha)$, $\tilde{C}(i,\alpha)$  and $\hat{C}(i,\alpha)$ for each choice of the weight. For each combination, vertical bars denote the range of the values, while the dot is the global coefficient for the whole network. Purple behavior is related to the coefficients computed on the unweighted network. In this case the three coefficients give the same results and do not depend on weights.
	}
	\label{ClosER}
\end{figure*}

We start focusing on ER graphs and we simulate a directed multiplex network adapting the logic of the classical $G(N,p)$ ER model to a multiplex framework. The graph is here constructed following a two-step algorithm.
First, we define the number of nodes $N$ and the number of layers $L$ and we connect nodes randomly. Each arc is included in the graph with attachment probability $p$, independently from every other arc. We use the same probability both for inter-layer and intra-layer connections. \\
Subsequently, let the weight of any existing arc to be drawn by an independent random variable in the interval $(0,1]$. In particular, we assume to simulate weights from a Beta distribution. We test different scenarios considering the same unweighted multilayer network and varying the average of the distribution of the weights from 0.1 to 0.9 with steps of 0.1. Therefore, we compare several networks with the same connections and different weights.
To this end, without loss of generality, in Figure \ref{ClustER} we consider a directed node-aligned multilayer network with 50 nodes and two layers and where arcs have been obtained assuming an attachment probability equal to 0.5. As expected, for the unweighted network the three alternative coefficients (see formulas (\ref{DeDomenico}), (\ref{ClementeGrassi}) and (\ref{Fagiolo})) provide the same results. Additionally, $C^{clust}$ is equal to the attachment probability as for monoplex network.\\
Considering the effect of weights, we observe that an increasing clustering is observed when average weight increases. This pattern can be explained by the fact that on average each observed triangle has a higher weight. It is noticeable how the alternative coefficients catch in a different way the effect of triangles. The coefficient $C(i,\alpha)$ tends to be more affected by the number of triangles detecting high clustered structures also when weights are lower. We have instead that the other two coefficients appear heavily affected by the weights. In particular, $\hat{C}(i,\alpha)$ provides coefficient close to zero when the average weight is equal to 0.1. This result occurs also if on average each node is involved in a half of the potential triangles (being $p=0.5$). The justification is related to the fact that the potential triangles have a weight equal to one, while the weight of actual triangles depends on the average weight observed in the network (see formula (\ref{DeDomenico})). \\
Figure 	\ref{ClosER} provides the analogous plot considering closure and clumping coefficients, respectively (see formulas (\ref{e:clos_nodo_livello}) and (\ref{e:clump_nodo_livello})). The behavior of these coefficients is very similar. We mainly notice a lower average value for closure with respect to clustering. This is mainly explained by the fact that although a high number of actual triangles are obtained, the increase of number of potential triangles is, in general, larger leading to a reduction of the coefficient. It is also observed a higher variability showing how the inclusion of different types of triangles emphasizes differences between nodes. Patterns of clumping are easily explained by the fact that this coefficient is a weighted average of clustering and closure coefficients.

\begin{figure*}
	\centering
	\includegraphics[width=12cm,height=8cm]{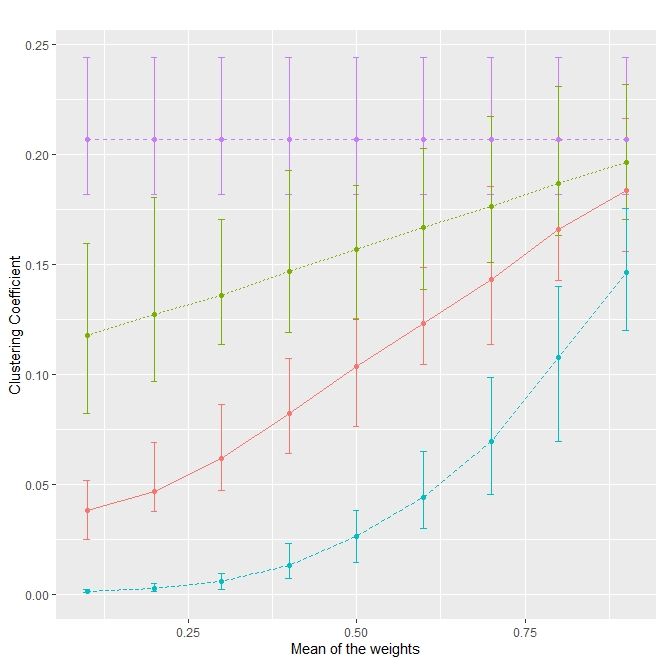}
	\caption{Comparison of clustering coefficients for a ER multilayer network, 50 nodes, 2 layers and attachment probability equal to 0.2. In green, red and light blue we report the distributions of $C(i,\alpha)$, $\tilde{C}(i,\alpha)$  and $\hat{C}(i,\alpha)$ for each choice of the weight. For each combination, vertical bars denote the range of the values, while the dot is the global coefficient for the whole network. Purple behavior is related to the coefficients computed on the unweighted network. In this case, the three coefficients give the same results and do not depend on weights.
	}
	\label{ClustER2}
\end{figure*}

While the number of nodes and layers has a limited impact on the average clustering, an interesting behavior is observed for different values of $p$. We report in Figure 	\ref{ClustER2}, the same comparison obtained considering a multilayer network where arcs are connected using an attachment probability equal to 0.2. As expected, a lower probability indicates a reduced number of triangles in the network, leading to a lower coefficient in the unweighted case. The introduction of weights produces a very similar average pattern to that observed in Figure \ref{ClustER}. However, a higher volatility between nodes is observed when the probability is reduced.  Since a lower density in observed, weights' volatility emphasizes differences between nodes.

\subsection{\added{Small-world graphs}}
\begin{figure*}
	\centering
	\includegraphics[width=12cm,height=8cm]{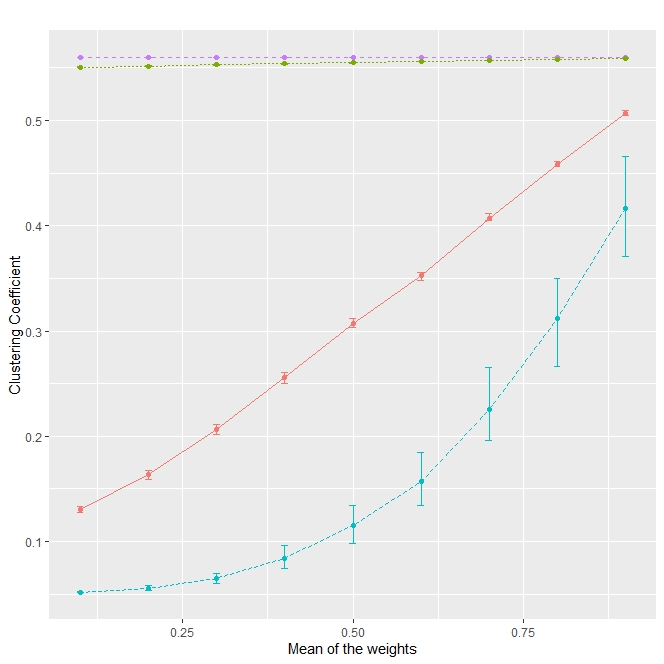}
	\caption{Comparison of clustering coefficients for a SW multilayer network, 50 nodes, 2 layers and rewiring probability equal to 0.5 and the same average degree of the ER model. In green, red and light blue we report the distributions of $C(i,\alpha)$, $\tilde{C}(i,\alpha)$  and $\hat{C}(i,\alpha)$ for each choice of the weight. For each combination, vertical bars denote the range of the values, while the dot is the global coefficient for the whole network. Purple behavior is related to the coefficients computed on the unweighted network. In this case, the three coefficients give the same results and do not depend on weights.
	}
	\label{ClustWS}
\end{figure*}

\begin{figure*}
	\centering
	\includegraphics[width=6cm,height=6cm]{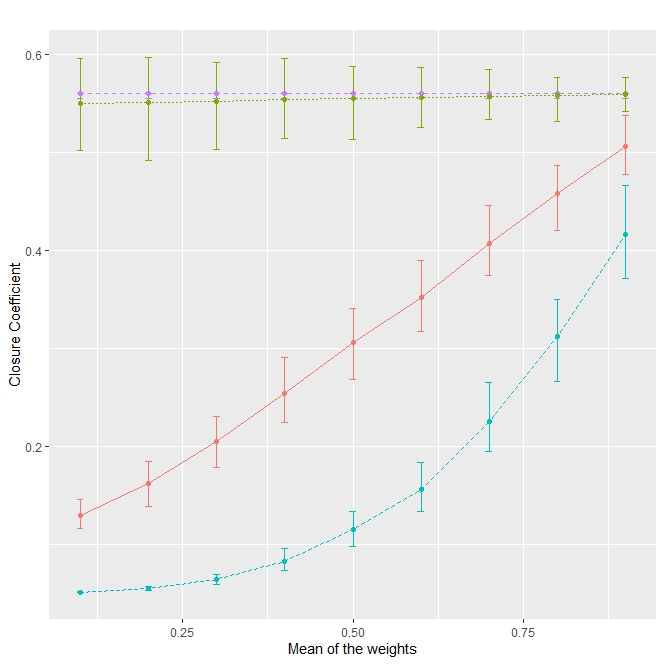}
	\includegraphics[width=6cm,height=6cm]{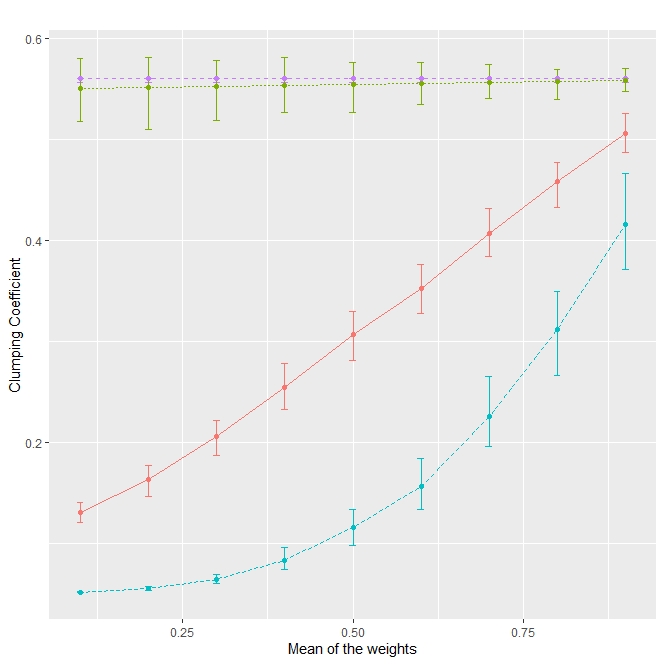}
	\caption{Comparison of closure and clumping coefficients for a SW multilayer network, 50 nodes, 2 layers and rewiring probability equal to 0.5. In green, red and light blue we report the distributions of $C(i,\alpha)$, $\tilde{C}(i,\alpha)$  and $\hat{C}(i,\alpha)$ for each choice of the weight. For each combination, vertical bars denote the range of the values, while the dot is the global coefficient for the whole network. Purple behavior is related to the coefficients computed on the unweighted network. In this case, the three coefficients give the same results and do not depend on weights.
	}
	\label{ClosWS}
\end{figure*}

To provide a comparison, we consider \added{in this subsection} a directed and weighted multilayer SW network. Traditionally, SW networks can be obtained, starting from a simulated lattice and rewiring each edge at random to a new target node, with probability $p$. As described in \cite{Watts1998},  a node is chosen and the arc that connects it to its nearest neighbor in a clockwise sense is considered. With probability $q$, this arc is reconnected to another node chosen uniformly at random over the entire ring, with duplicate arcs in the same direction forbidden; otherwise the arc is left in place. The process is repeated by moving clockwise around the ring, considering each vertex in turn until one lap is completed. Next, the arcs that connect nodes to their second-nearest neighbors clockwise are considered. As before, each of these arcs is randomly rewired with probability $q$ continuing the process, circulating around the ring and proceeding outward to more distant neighbors after each lap, until each arc in the original lattice has been considered once. In this case, we generate with this procedure a network with a number of nodes equal to  $N*L$ and then we transform the network in a multilayer network with $N$ nodes and $L$ layers. As for the ER graph, weights are then added in a second step by randomly assigning to each existing arc a value simulated from a Beta distribution. \\
We report in Figure \ref{ClustWS} main results obtained considering a rewiring probability equal to $0.5$ and the same average degree of the ER model analyzed above. As well known, intermediate values of $q$ result in a small-world network that shares properties of either regular or random graphs. Indeed, connected SW networks trivially have small average path lengths and high clustering coefficients. However, typically for these values of $q$ smaller differences are noticed with respect to random graphs. This result is confirmed also in this multilayer setting. We have indeed that the unweighted coefficient is slightly higher (approximately 0.56) than the corresponding coefficient displayed in Figure \ref{ClustER} for an analogous ER network. We observe that also the introduction of weights confirms the higher level of interconnections observed in this network. Although the patterns of different clustering coefficients are very similar to that of Figure \ref{ClustER}, we notice an average increase of values for all the combinations. Also closure and clumping coefficients show higher values for the SW network (See Figure \ref{ClosWS})

\begin{figure*}
	\centering
	\includegraphics[width=6cm,height=6cm]{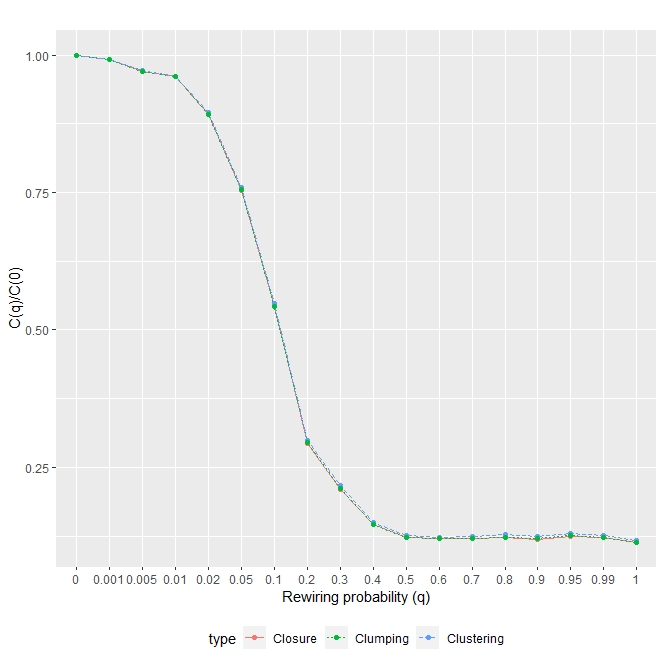}
	\includegraphics[width=6cm,height=6cm]{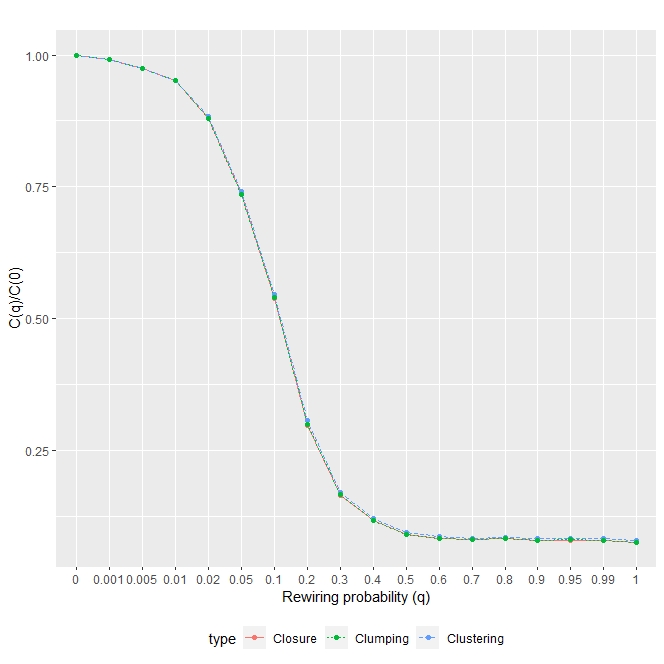}
	\caption{We considered several unweighted SW multilayer networks with 100 nodes, 2 layers (left side) and 3 layers (right side) and average degree equal to 10. Each network has been obtained varying the rewiring probability $q \in (0,1]$. Each dot is the ratio between global clustering coefficient computed on the SW network with probability $q$ and the clustering coefficient of the regular network. The same ratios are computed also using closure and clumping coefficients.
	}
	\label{VP}
\end{figure*}

To exploit the behavior of the clustering coefficient on an unweighted SW network, we analyzed the patterns between complete regularity ($q = 0$) and total disorder
($q = 1$), studying the intermediate region ($0 < q < 1$). In particular, in Figure \ref{VP} (left side), we consider an unweighted directed multilayer network with 100 nodes, 2 layers and average degree equal to 10. We display the ratio $\frac{C(q)}{C(0)}$ between the global clustering coefficient of SW network obtained for specific values of the rewiring probability $q$ and the global clustering coefficient of a regular network (i.e. $C(0)$). As for the monoplex case (see, e.g., \cite{Watts1998}, \cite{Humpries}, \cite{CleFGra}) it is confirmed that there is a broad interval of $q$ over which clustering coefficient is significantly greater than
random. These small-world networks result from the immediate drop in the average path caused by the introduction of a few long-range edges, but they keep a very high clustering.  It is also interesting to note that a very similar behavior is observed for closure and, therefore, for clumping. \\
In Figure \ref{VP} (right side), we repeated the analysis increasing the number of layers and, although lower coefficients are observed, the same pattern can be noticed when the probability $q$ varies.

\section{Conclusions}
\label{Conclusions}
In this paper we extend to the context of weighted multilayer networks the clustering and closure coefficients through the introduction of the clumping coefficient, that generalizes the previous ones considering incomplete triangles of any type. 
Our proposal overcomes the fictitious distinction between the two coefficients based on a conventional choice in completing open triads of nodes. \\
Moreover, the introduction of this new coefficient allows us to organize the existing definitions in a systematic taxonomy. In particular, we show how coefficients in the literature descend as particular cases from a more general definition.\\
We also adapt these coefficients to the different scales that characterizes a multilayer network, in order to study their structure from different perspectives, according to the point of view of interest on the network. 
In particular, local coefficients are introduced for a single node on a single level, together with coefficients that capture the mesoscale structure of all the replicas of the same node on all levels or of all nodes within a single level. A global coefficient is then introduced in such a way as to reproduce the classical idea of transitivity for single layer networks.
All introduced coefficients are then specified for different classes of directed triangles, namely out, in, cycle and middleman.\\ 
We formally represent all the proposed coefficients using the tensor formalism. This allows us to unify all the definitions existing in the literature together with the new ones here introduced. 
With a numerical experiment, we apply all the proposed indicators to random multilayer networks, built according to specific algorithms and with a suitable weights distribution. We show the effectiveness of cohesion coefficients in capturing different peculiarities of the network structure, at different scales, and we perform a sensitivity analysis of their dependence on some key parameters.
Concluding, usually in real multilayer networks the nature of the links between levels is different from that of the links within a level. This often makes meaningful an extension of the definition of a triangle to the case in which the jump from a level to another is not counted as a side of the triangle itself but is considered as an arc without cost. This observation opens up the possibility of generalized triangles with more than three sides.  
In the approach presented in this paper, the extension to this type of triangles could be immediate, by introducing separate adjacency tensors for the arcs within the levels and the arcs between levels and building up the corresponding tensor products to calculate the number of triangles of interest. We deal with this issue in a future research.

\vskip 3truecm

\bibliographystyle{elsarticle-num}


\newpage

\appendix

\section{\added{Detailed tables}}

In this Appendix we collect Tables containing the number of actual and potential triangles of various types (in, out, cycle, middleman and total), referred to the node $i$ on level $a$ (Tables \ref{TableA1}-\ref{TableA3}). Table \ref{TableA4} collects the total clustering, closure and clumping coefficients according to the different versions discussed in Section \ref{Clustering coefficients for weighted DMN}. Table \ref{TableA5} exemplifies the out coefficients. Similar tables can be provided for the in, cycle and middleman coefficients. \\
To simplify the reading, tensors appear in Tables following the notations in \eqref{definition} and \eqref{definition2} and setting
$H_1 H_2 H_3:= [{H_1}]^{\mu \alpha}_{\nu \beta}[{H_2}]^{\nu \beta}_{\rho \gamma}[{H_3}]^{\rho \gamma}_{\sigma \delta}E_{\mu \alpha}(i,a)E^{\sigma \delta}(i,a)$.

\begin{table*}[!h]
	\centering{}
	\renewcommand{\arraystretch}{1.2}
	\begin{tabular}{|l|c|c|}
		\hline \hline
		\multicolumn{3}{|c|}{\bf Actual triangles}\tabularnewline
		\hline 
		{\bf Class} & {\bf Notation} & {\bf Formula} \tabularnewline
		\hline
		Out-triangles & $T_{\rm out}^{\rm act}(i,a)$  & ${A}_{\rm out} {A}_{\rm out} {A}_{\rm in}={A}_{\rm out} {A}_{\rm in} {A}_{\rm in}$  \tabularnewline \hline
		In-triangles & $T_{\rm in}^{\rm act}(i,a)$  & ${A}_{\rm in} {A}_{\rm out} {A}_{\rm out}={A}_{\rm in} {A}_{\rm in} {A}_{\rm out}$  \tabularnewline \hline
		Cycle-triangles & $T_{\rm cycle}^{\rm act}(i,a)$  & ${A}_{\rm out} {A}_{\rm out} {A}_{\rm out}={A}_{\rm in}{A}_{\rm in}{A}_{\rm in}$  \tabularnewline \hline
		Middleman-triangles & $T_{\rm mid}^{\rm act}(i,a)$  & ${A}_{\rm out} {A}_{\rm in} {A}_{\rm out}={A}_{\rm in}{A}_{\rm out}{A}_{\rm in}$  \tabularnewline \hline
		Total & $T_{\rm tot}^{\rm act}(i,a)$  & $4{\tilde A}{\tilde A}{\tilde A}$  \tabularnewline \hline
		\hline
	\end{tabular}\caption{Number of actual triangles for the node $i$ on layer $a$}
	\label{TableA1}
\end{table*}

\begin{table*}[!h]
	\centering{}
	\renewcommand{\arraystretch}{1.2}
	\begin{tabular}{|l|c||c|}
		\hline \hline
		\multicolumn{3}{|c|}{\bf Potential triangles of the first type - I}\tabularnewline
		\hline 
		{\bf Class} & {\bf Notation} & {\bf Formula} \tabularnewline
		\hline
		Out-triangles & $T_{\rm out}^{\rm I}(i,a)$ & ${A}_{\rm out} F {A}_{\rm in}$  \tabularnewline \hline
		In-triangles & $T_{\rm in}^{\rm I}(i,a)$ & ${A}_{\rm in} F {A}_{\rm out}$  \tabularnewline \hline
		Cycle-triangles &$T_{\rm cycle}^{\rm I}(i,a)$ & ${A}_{\rm out} F {A}_{\rm out}={A}_{\rm in} F {A}_{\rm in}$  \tabularnewline \hline
		Middleman-triangles &$T_{\rm mid}^{\rm I}(i,a)$ & ${A}_{\rm out} F {A}_{\rm out}={A}_{\rm in} F {A}_{\rm in}$ \tabularnewline \hline
		Total &$T_{\rm tot}^{\rm I}(i,a)$ & $4{\tilde A}F{\tilde A}$  \tabularnewline \hline
		\hline
	\end{tabular}\caption{Number of potential triangles when the focal node is the central one in the open triad}
	\label{TableA2}
\end{table*}

\begin{table*}[!h]
	\centering{}
	\renewcommand{\arraystretch}{1.2}
	\begin{tabular}{|l|c||c|}
		\hline \hline
		\multicolumn{3}{|c|}{\bf Potential triangles of the second type - II}\tabularnewline
		\hline 
		{\bf Class}  & {\bf Notation}& {\bf Formula} \tabularnewline
		\hline
		Out-triangles  & $T_{\rm out}^{\rm II}(i,a)$ & $F{A}_{\rm out}{A}_{\rm in}+{A}_{\rm out}{A}_{\rm out}F$  \tabularnewline \hline
		In-triangles  & $T_{\rm in}^{\rm II}(i,a)$ & $F{A}_{\rm in}{A}_{\rm out}+{A}_{\rm in}{A}_{\rm in}F$   \tabularnewline \hline
		Cycle-triangles &$T_{\rm cycle}^{\rm II}(i,a)$ & $F{A}_{\rm out}{A}_{\rm out}+{A}_{\rm out}{A}_{\rm out}F$  \tabularnewline \hline
		Middleman-triangles &$T_{\rm mid}^{\rm II}(i,a)$ & $F{A}_{\rm in}{A}_{\rm out}+{A}_{\rm out}{A}_{\rm in}F$ \tabularnewline \hline
		Total &$T_{\rm tot}^{\rm II}(i,a)$ & $4\big( F{\tilde A}{\tilde A} + {\tilde A}{\tilde A}F\big)$  \tabularnewline \hline
		\hline
	\end{tabular}\caption{Number of potential triangles when the focal node is one of the ends in the open triad.}
	\label{TableA3}
\end{table*}

\begin{table*}[!h]
	\small
	\centering{}
	\begin{tabular}{|c||c|c|c|}
		\hline \hline
		\multicolumn{4}{|c|}{\bf Coefficients for weighted DMN}\tabularnewline
		\hline 
		
		Coefficient		& Clustering & Closure & Clumping  \tabularnewline
		\hline 
		\centering
		$\tilde{C}(i,a)$ & $\frac{\tilde{M} \tilde{M} \tilde{M}}{\tilde{M} F \tilde{M}}$ & $\frac{2\tilde{M} \tilde{M} \tilde{M}}{F\tilde{M} \tilde{M}+\tilde{M} \tilde{M}F}$ & $\frac{3\tilde{M} \tilde{M} \tilde{M}}{F\tilde{M} \tilde{M}+\tilde{M} F \tilde{M}+\tilde{M} \tilde{M}F}$   \tabularnewline
		${C(i,a)}$ & $\frac{\tilde{M} \tilde{A} \tilde{A}}{\tilde{M} F \tilde{A}}$ & $\frac{2\tilde{M} \tilde{M} \tilde{M}}{F\tilde{A} \tilde{M}+\tilde{A} \tilde{M}F}$  & $\frac{3\tilde{M} \tilde{M} \tilde{M}}{F\tilde{A} \tilde{M}+\tilde{M} F \tilde{A}+\tilde{A} \tilde{M}F}$   \tabularnewline
		$\hat{C}(i,a)$  & $\frac{\hat{M} \hat{M} \hat{M}}{\hat{A} F \hat{A}}$ & $\frac{2\hat{M} \hat{M} \hat{M}}{F\hat{A} \hat{A}+\hat{A} \hat{A}F}$  & $\frac{3\hat{M} \hat{M} \hat{M}}{F\hat{A} \hat{A}+\hat{A} F \hat{A}+\hat{A}\hat{A}F}$   \tabularnewline
		\hline \hline
	\end{tabular}
	\caption{Overview of the local coefficients for a weighted DMN considering total triangles}
	\label{TableA4}
\end{table*}

\begin{table*}
	\small
	\centering{}
	\begin{tabular}{|c||c|c|c|}
		\hline \hline
		\multicolumn{4}{|c|}{\bf Out coefficients for weighted DMN}\tabularnewline
		\hline 
		
		Coefficient		& Clustering & Closure & Clumping  \tabularnewline
		\hline 
		\centering
		$\tilde{C}(i,a)$ & $\frac{\tilde{M}_{\rm out} \tilde{M}_{\rm out} \tilde{M}_{\rm in}}{\tilde{M}_{\rm out} F \tilde{M}_{\rm in}}$ & $\frac{2\tilde{M}_{\rm out} \tilde{M}_{\rm out} \tilde{M}_{\rm in}}{F\tilde{M}_{\rm out} \tilde{M}_{\rm in}+\tilde{M}_{\rm out} \tilde{M}_{\rm out}F}$ & $\frac{3\tilde{M}_{\rm out} \tilde{M}_{\rm out} \tilde{M}_{\rm in}}{F\tilde{M}_{\rm out} \tilde{M}_{\rm in}+\tilde{M}_{\rm out} F \tilde{M}_{\rm in}+\tilde{M}_{\rm out} \tilde{M}_{\rm out}F}$   \tabularnewline
		${C(i,a)}$ & $\frac{\tilde{M}_{\rm out} \tilde{A}_{\rm out} \tilde{A}_{\rm in}}{\tilde{M}_{\rm out} F \tilde{A}_{\rm  in}}$ & $\frac{2\tilde{M}_{\rm out} \tilde{M}_{\rm out} \tilde{M}_{\rm in}}{F\tilde{A}_{\rm out} \tilde{M}_{\rm in}+\tilde{A}_{\rm out} \tilde{M}_{\rm out}F}$  & $\frac{3\tilde{M}_{\rm out} \tilde{M}_{\rm out} \tilde{M}_{\rm in}}{F\tilde{A}_{\rm out} \tilde{M}_{\rm in}+\tilde{M}_{\rm out} F \tilde{A}_{\rm in}+\tilde{A}_{\rm out} \tilde{M}_{\rm out}F}$   \tabularnewline
		$\hat{C}(i,a)$  & $\frac{\hat{M}_{\rm out} \hat{M}_{\rm out} \hat{M}_{\rm in}}{\hat{A}_{\rm out} F \hat{A}_{\rm in}}$ & $\frac{2\hat{M}_{\rm out} \hat{M}_{\rm out} \hat{M}_{\rm in}}{F\hat{A}_{\rm out} \hat{A}_{\rm in}+\hat{A}_{\rm out} \hat{A}_{\rm out}F}$  & $\frac{3\hat{M}_{\rm out} \hat{M}_{\rm out} \hat{M}_{\rm in}}{F\hat{A}_{\rm out} \hat{A}_{\rm in}+\hat{A}_{\rm out} F \hat{A}_{\rm in}+\hat{A}_{\rm out}\hat{A}_{\rm out}F}$   \tabularnewline
		\hline \hline
	\end{tabular}
	\caption{Overview of the local out coefficients for weighted DMN}
	\label{TableA5}
\end{table*}

\end{document}